\documentstyle[a4,11pt,epsf,epsfig,twoside]{article}
\begin{document}
\setcounter{page}{1}
\title{R$\&$D ON THE GEM READOUT OF THE TESLA TPC}
\author{M. HAMANN \\ {\it DESY Hamburg, Germany} 
}
\date{}
\maketitle
\begin{abstract}
\noindent
Studies for the TESLA TPC (Time Projection Chamber) 
with GEM (Gas Electron Multiplier) readout at 
DESY/Hamburg University are presented. 
Two test chamber setups are being operated.
The studies include basic GEM
performance, tracking and the determination of the resolution using
different pad sizes and geometries.
Our measurements show that chevron shaped pads lead 
to a better point resolution compared to rectangles. 
A second focus of our R$\&$D activities is the measurement of the 
ion feedback. It is determined to be in the order of a few percent
using a double GEM structure.

\end{abstract}

\section{Introduction}
A Time Projection Chamber (TPC, \cite{tpc}) is proposed to 
be the main tracking
device for the detector at the TESLA collider \cite{tdr}.\\
Our purpose is to study the TPC readout using Gas Electron Multipliers
(GEMs,  \cite{sauli97}) instead of conventionally used 
wire chambers to produce the gas amplification.
GEMs offer several advantages:
$\vec{E} \times \vec{B}$ 
effects degrading the spatial resolution are strongly reduced, and   
the amount of required material in the endcap is low;
ion feedback into the drift volume leading to field distortion 
is naturally suppressed to a level of a few percent. Due to
the absence of ion tails in the pulse shape the use of GEMs 
improves the intrinsic two track resolution in the drift direction 
\cite{sauli99}. \\
In the following our present TPC test setups and the measurements are
presented.

\section{Setups and Measurements}

\subsection{Large prototype}

\begin{figure}[htbp]
\begin{center}
\unitlength1.0cm
\begin{picture}(8,3.5) 
\put(-2.2,0.0){\epsfig{file=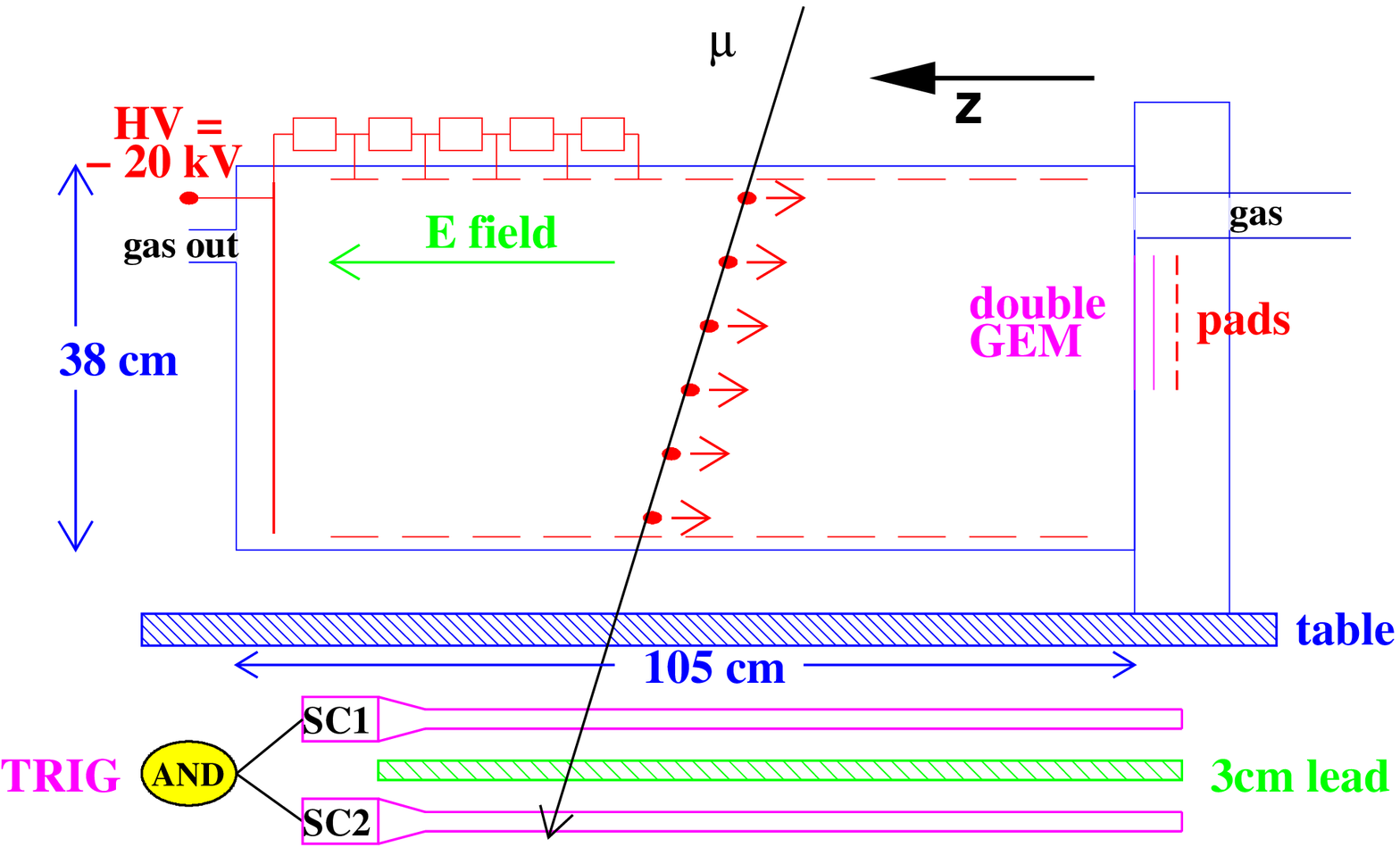,width=7.0cm}}
\put(6.,3.4){\epsfig{file=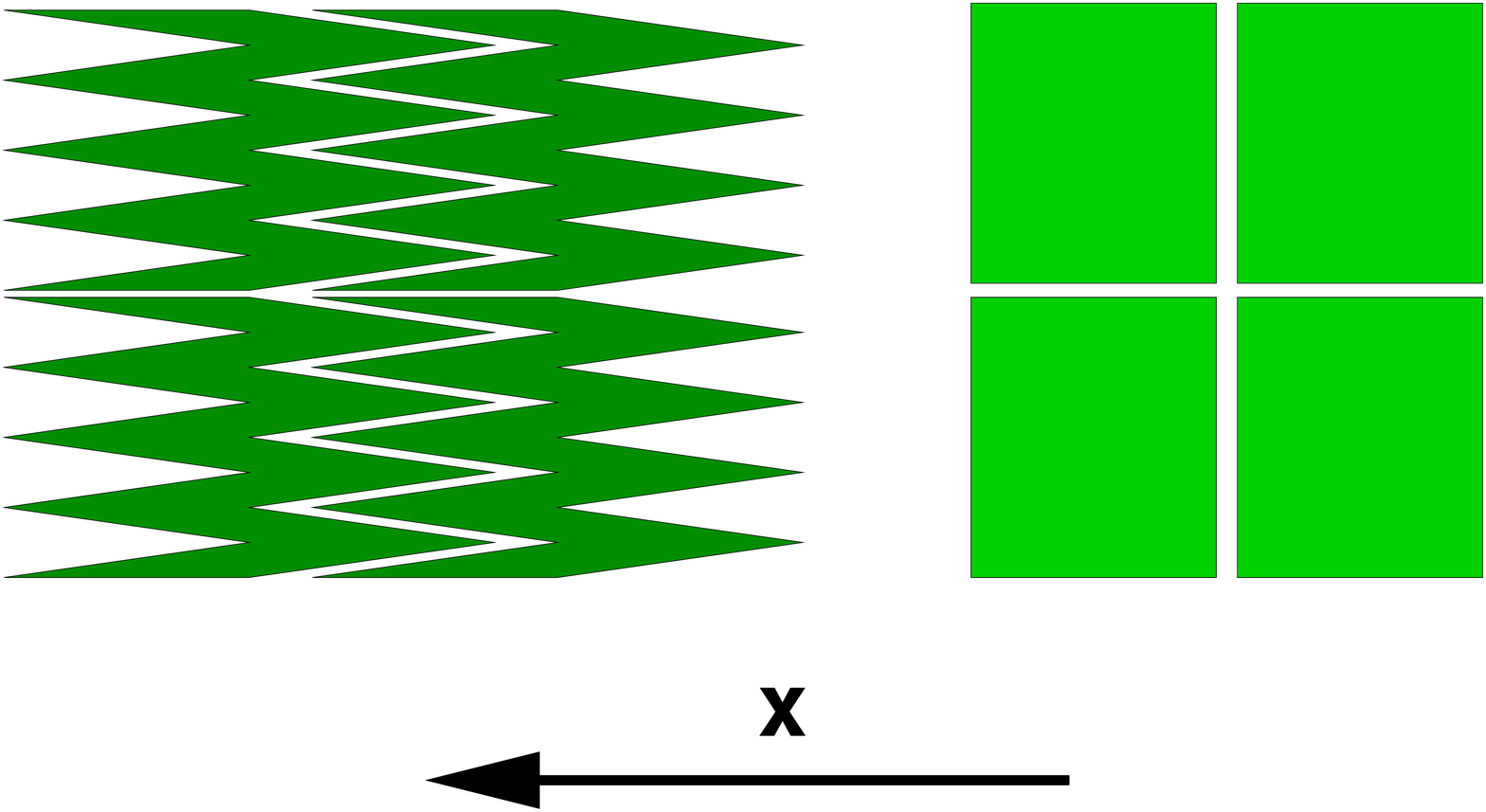,width=4.5cm,angle=180}}
\end{picture}
\caption[Schematic picture of the setup and pad geometries ]{\it
  Schematic picture of the setup and pad geometries}
\label{setup1}
\end{center}
\end{figure}

{\bf The setup} is shown in figure \ref{setup1}, left side.
It consists of the field cage, the endplate with the GEM module and 
two scintillators forming the trigger system \cite{lcntpc}. The
readout is done using an $11.4 \,{\mathrm MHz}$ Flash ADC.
The chamber volume is filled with a gas mixture of  
$93\,\%$ argon, $2\,\%$ carbon dioxide and $5\,\%$ methane (TESLA TDR \cite{tdr}).
Cosmic muons are used as ionising particles. No magnetic field is
applied ($B=0\,{\mathrm T}$).

\begin{figure}[htbp]
\begin{center}
\unitlength1.0cm
\begin{picture}(8.0,6.1) 
\put(-3.5,3.){\epsfig{file=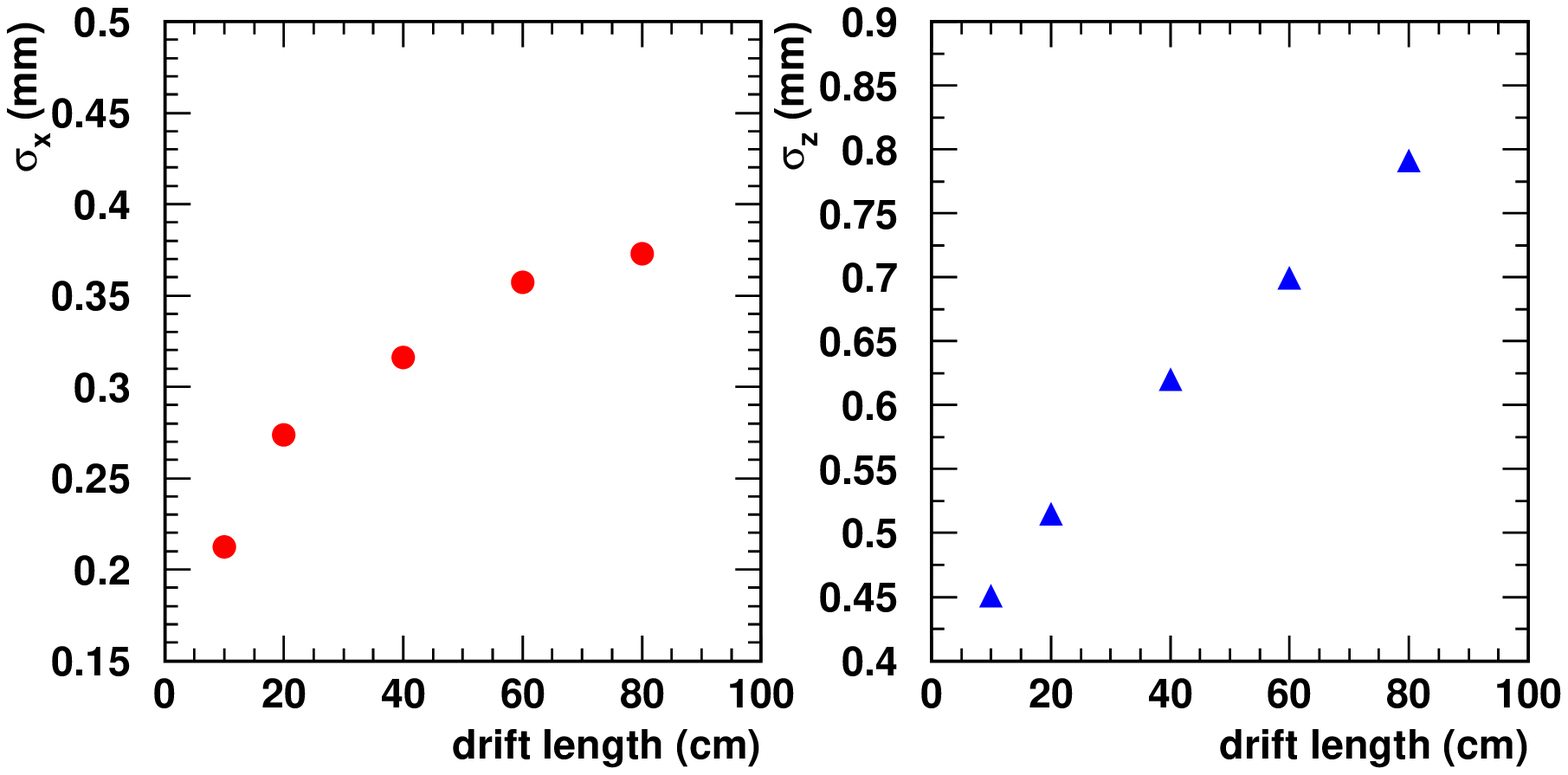,width=7.5cm}}
\put(-3.3,-0.3){\epsfig{file=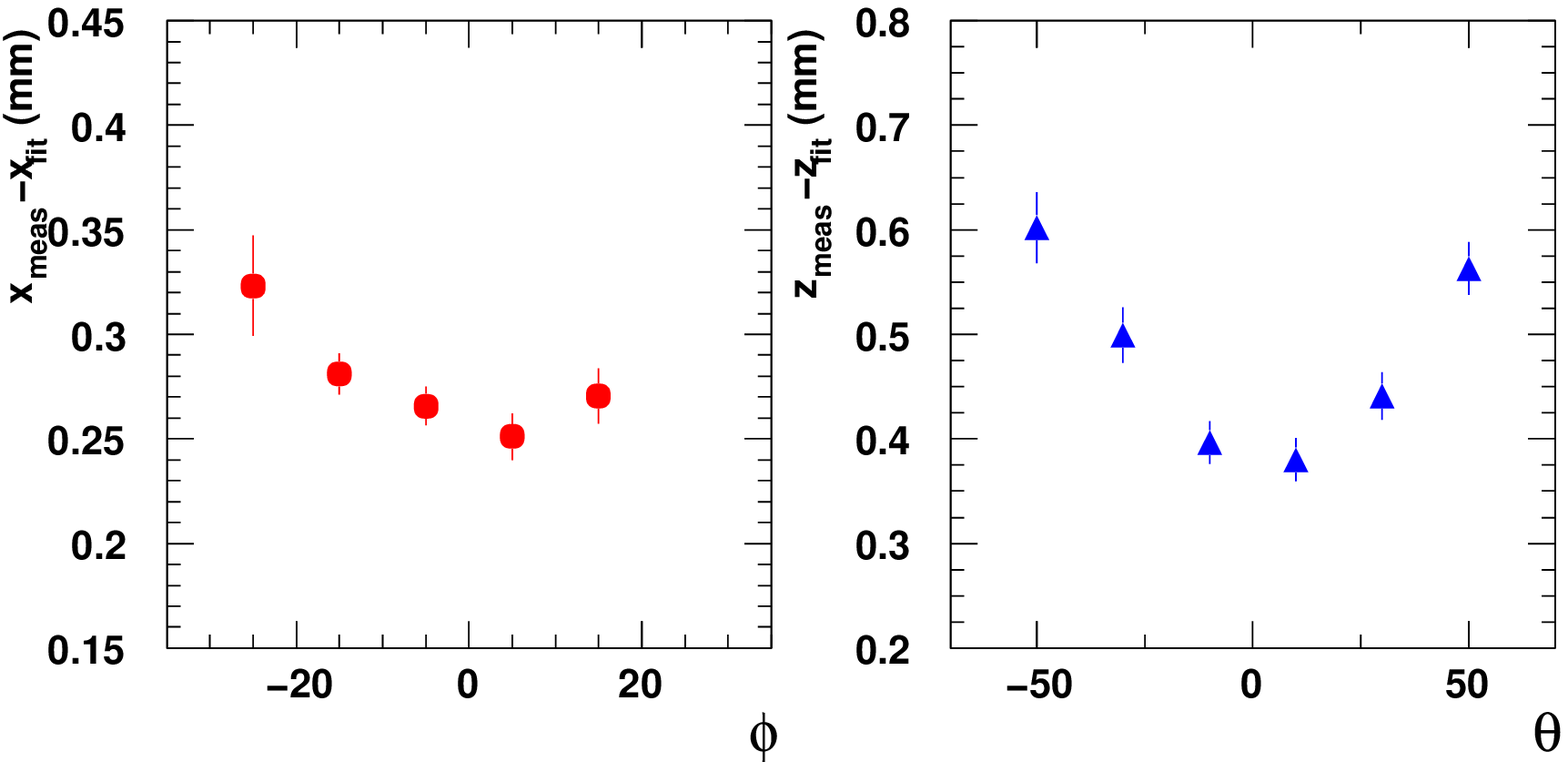,width=6.8cm}}
\put(4.3,-0.5){\epsfig{file=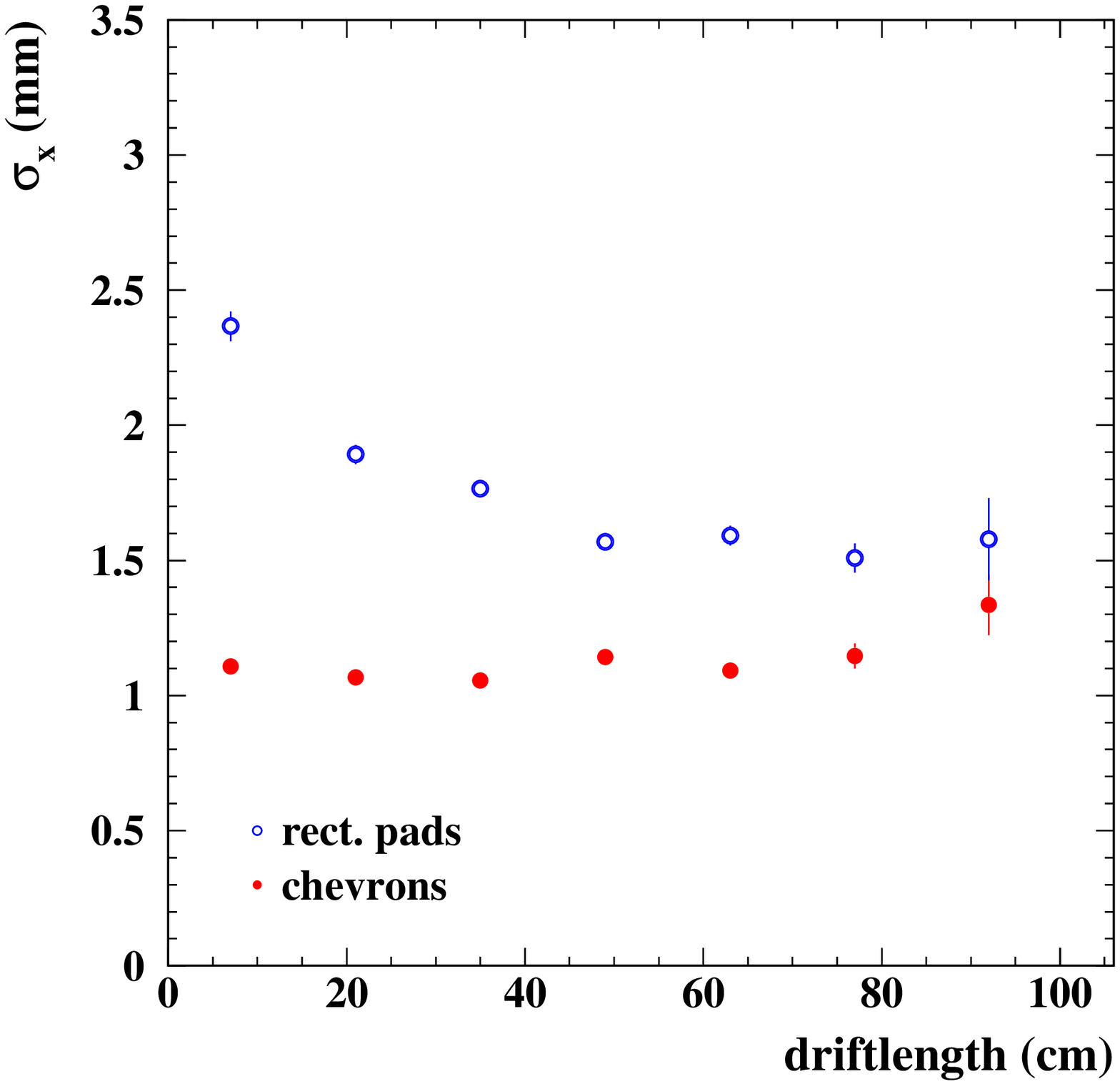,width=7.5cm}}
\end{picture}
\caption[Resolution]{\it Resolution of small pads (left) and large
  pads (right), see text}
\label{res1}
\end{center}
\end{figure}

\noindent
The endplate contains the GEM module consisting of two 
"standard" GEMs \cite{bach99}: A copper coated  capton foil
with holes of a double conical shape (inner(outer) diameter: $55(70) \,
\mu{\mathrm m}$) and a pitch of $140 \, \mu{\mathrm m}$. 
Each foil has a size of $10\,\times \, 10\,{\mathrm cm}^{2}$.
They are mounted in cascade with a gap
of $1.5\,{\mathrm mm}$ between the two
GEMs (transfer gap) and between the GEM and the array of readout pads,
respectively (induction gap). The corresponding fields are 
$E_{\mathrm{induction}} = E_{\mathrm{transfer}} \approx  1.3 \,\mathrm{kV/cm}$.\\
Our standard readout pads are rectangles 
of a size of $2\times 6 \,\mathrm{mm}^2$, as proposed in the TESLA
TDR. According to simulations \cite{schum01} more sophisticated 
pad geometries may lead to a better point resolution due to larger
charge sharing. 
Therefore we also performed measurements using chevron shaped pads.
Since the studies are made without magnetic field, the pad sizes are
scaled in $x$ according to the width of the charge cloud, governed by the
diffusion coefficient ${\mathcal D} (B)$. 
In the drift volume, ${\mathcal D}(B=0\,{\mathrm T})\approx 6\cdot 
{\mathcal D}(B_{\mathrm TESLA}=4\,{\mathrm T})$ holds true for our gas. 
Thus, we use rectangles with an extension in 
$x$ of $14 \,\mathrm{mm}$ and the corresponding chevron size 
(fig. \ref{setup1}, right).\\
{\bf Measurements:} Some already intensively studied GEM properties
\cite{bach99} are verified for a TPC: The gas
amplification depends exponentially on the 
applied voltages across the GEMs and the stability of the gain is
satisfactory: The gain variation equals
$1.9\%$ over a period of 70 hours. Taking into account the correction
for atmospheric pressure, we expect to reduce
the variation to a value  $< 1\, \%$, which is needed to pin down 
the relative error $\frac{\sigma_{{\mathrm d}E/{\mathrm
  d}x}}{{\mathrm d}E/{\mathrm d}x}$ below $5\, \%$.
The point resolution $\sigma_x$ (respectively  $\sigma_z$) 
is determined calculating the residuals $x_{\mathrm meas}-x_{\mathrm
  fit}$, $x_{\mathrm meas}$ being the centre of charge of the 3 d
electron cluster assuming a linear charge distribution and
$x_{\mathrm fit}$ the coordinate calculated from the linear track
fit, see figure \ref{res1}.  The four left plots contain results of the 
small rectangular pads: $x$ and $z$ resolution as a function of the drift
length (upper plots): Apart from very short drift distances, 
the charge cloud is larger than the pad size, and the
resolution is dominated by diffusion. In $x$, this leads 
to an overall worse result than expected for TESLA
($\sigma_{x,{\mathrm TESLA}} \approx 140 \, {\mathrm \mu m}$).  
The lower two plots show the degradation of the resolution 
with increasing track angle, due to the prolongation of the projection
of the charge cloud on the corresponding coordinate ($x$ and $z$). 
The right plot in figure \ref{res1}
contains the results using large pads. The resolution $\sigma_x$ 
is dominated by the pad size rather than by diffusion.
For small drift distances the point resolution with chevrons
is better than with rectangles. This effect of the pad geometry 
gets smaller with increasing drift distance. 
The corresponding resolution at TESLA can be estimated by scaling down 
the measured resolution by the factor corresponding to the ratio of
our pad sizes.
\subsection{Small prototype ("Mini TPC")}
\begin{figure}[htbp]
\begin{center}
\unitlength1.0cm
\begin{picture}(12.,4.3) 
\put(-1.5,0.0){\epsfig{file=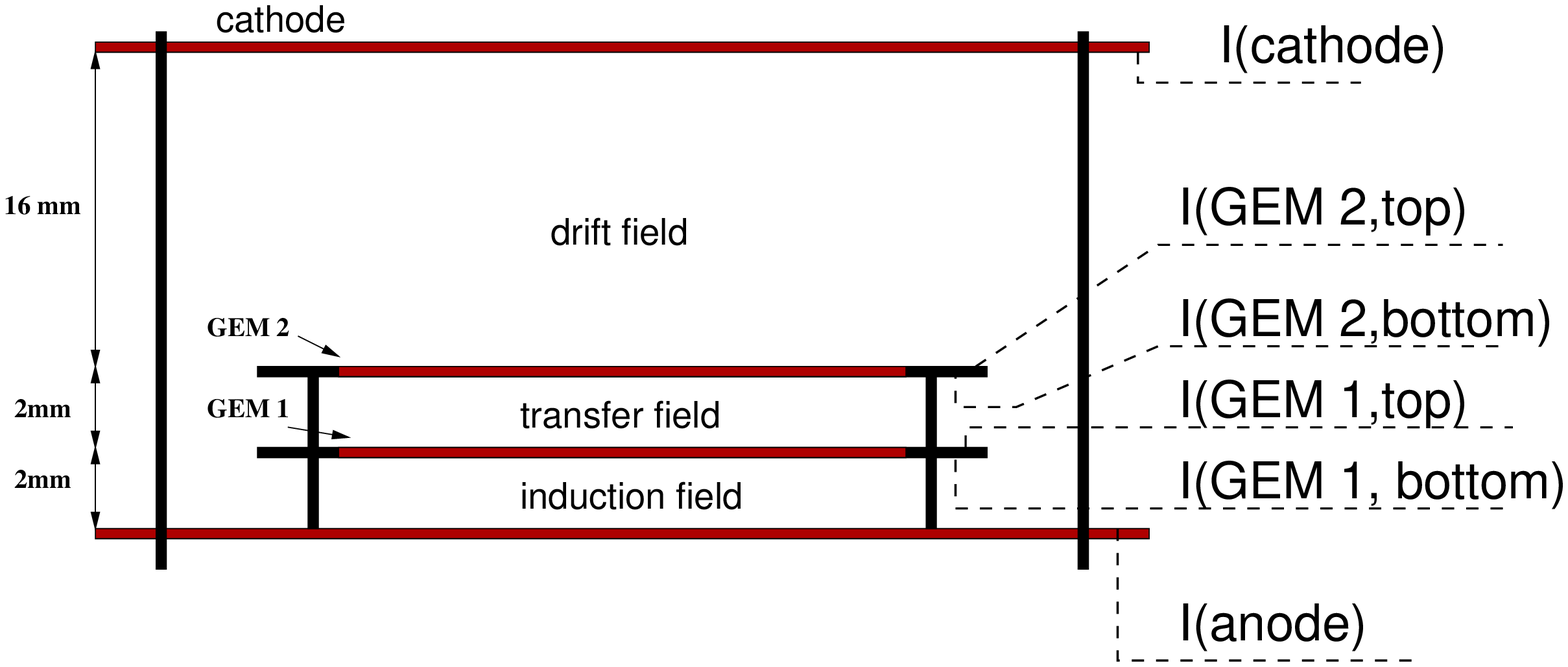,width=8.5cm}}
\put(8.,-0.7){\epsfig{file=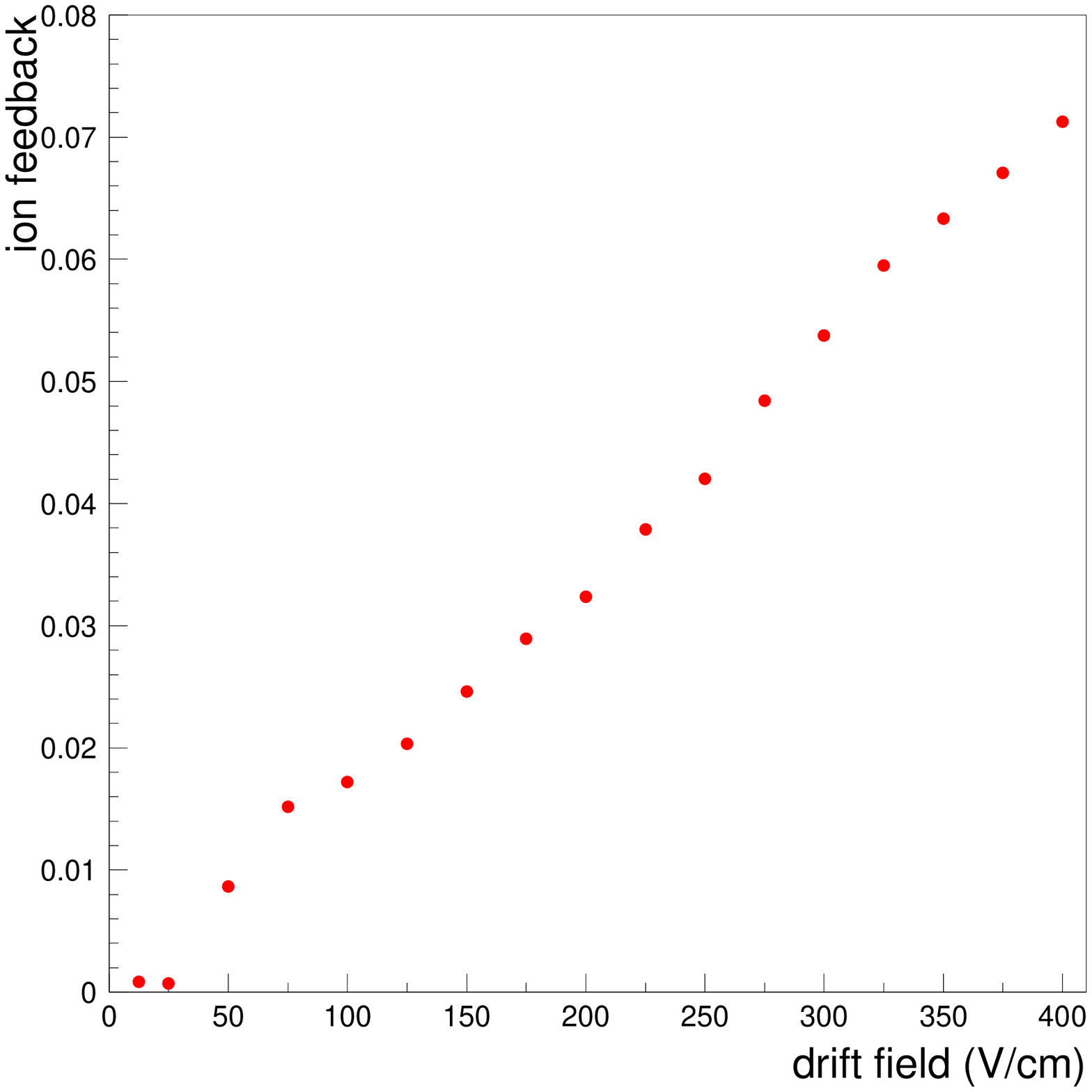,width=6cm}}
\end{picture}
\caption[Ion feedback measurement: "Mini TPC"]{\it Ion feedback
  measurement: "Mini TPC", see text}
\label{setup2}
\end{center}
\end{figure}

A side view sketch of the "mini TPC" \cite{lux01} can be seen in  figure \ref{setup2}. 
It is built to measure the ion feedback and  
features a short drift distance ($16 \,{\mathrm mm}$) making it possible to run at a 
moderate cathode voltage ($U_{\mathrm cath} \approx  1.5 \, {\mathrm kV}$) which allows
the measurement of the cathode current $I_{\mathrm cath}$
(in the order of few ${\mathrm nA}$). 
Dividing $I_{\mathrm cath}$ by the sum of all positive currents gives
the ion feedback \cite{bach99}. 
The used gas mixture is the same as in the previous case (TDR gas), 
the ionisation  is produced by a radioactive iron source.
In the plot of figure \ref{setup2} the ion feedback is shown as a 
function of the drift field. Both induction and transfer field are  
adjusted to $1\,{\mathrm kV/cm}$, the GEM voltages are $U_{\mathrm GEM 2}=390 \,{\mathrm V}$
and $U_{\mathrm GEM 1}=370 \,{\mathrm V}$.
For a drift field of $\approx 250\,{\mathrm V/m}$, as
foreseen for the TESLA detector, the feedback is $\approx 4\,\%$.

\section{Future Plans}
The future measurement programme includes GEM operation 
in a magnetic field up to $5 \,{\mathrm T}$
and TPC studies in the  DESY electron test beam ($6\,{\mathrm GeV}$).
The determination of the point resolution under more realistic 
conditions will be possible and it is planned to
study the two track resolution. 
Resolution studies will also be carried out with different 
GEM tower geometries. 
Simulations will accompany the measurement programme.

\section{Summary and Conclusion}
Studies on the GEM readout for the TESLA TPC are performed. 
The gain variation without correction for atmospheric pressure 
is lower than $2\, \%$ which 
is promising for ${\mathrm d}E/{\mathrm d}x$ measurements.
Resolution studies in $x$ with $B=0\,{\mathrm T}$ show different behaviour 
of rectangular and chevron pads favouring chevrons especially for 
short drift distances. The $z$ resolution is $0.5 - 1.0 \, {\mathrm mm}$.
It has to be taken into account that resolution measurements
depend strongly on the diffusion --- thus on the gas mixture and on
the geometry of the GEM tower.
Finally, for a double GEM structure, 
the ion feedback is determined to be in the order of a
few percent.

\section*{Acknowledgement}

I would like to thank T. Behnke, N. Ghodbane, T. Kuhl, T. Lux and F. Sefkow 
for their support and contributions.

\end{document}